\documentclass[aps,pra,floatfix,twocolumn,nofootinbib,showpacs]{revtex4}

\usepackage{amsmath}
\usepackage{amssymb}
\usepackage{graphicx}
\usepackage{dcolumn}
\usepackage[sort&compress]{natbib}
\usepackage{bm}
\usepackage[dvips]{epsfig}

\usepackage{subfigure} 
\usepackage{float}

\newcommand{\comment}[1]{}
\newcommand{\BEQ}{\begin{equation}}
\newcommand{\EEQ}{\end{equation}}
\newcommand{\BEA}{\begin{eqnarray}}
\newcommand{\EEA}{\end{eqnarray}}
\newcommand{\ssum}{{\sum}}

\renewcommand{\a}{\alpha}
\renewcommand{\b}{\beta}
\newcommand{\yt}{\tilde{y}}

\begin{document}

\title{Coevolutionary networks of reinforcement-learning agents}

\author{ Ardeshir Kianercy and  Aram Galstyan}
\affiliation{Information Sciences Institute, University of Southern California, Marina del Rey, CA 90292, USA}

\begin{abstract}
This paper presents a  model of network formation in repeated games where the players  adapt their strategies and network ties simultaneously using a simple reinforcement learning scheme. It is demonstrated that the co-evolutionary dynamics of such systems can be described via coupled replicator equations.  We provide a comprehensive analysis  for three-player two-action games, which is the minimum  system size with nontrivial structural dynamics. In particular, we characterize the Nash equilibria (NE) in such games and examine the local stability of the rest points corresponding to those equilibria. We also study general $n$-player networks via both simulations and analytical methods and find that in the absence of exploration, the  stable equilibria  consist of {\em{star}} motifs  as the main building blocks of the network. Furthermore, in  all stable equilibria the agents play pure strategies, even when the game allows mixed NE. Finally, we study the impact of exploration on learning outcomes, and observe that there is a critical exploration rate above which the symmetric and uniformly connected network topology becomes stable. 
  
  \end{abstract}

\maketitle
\section{Introduction} 
\label{sec:intro}

Networks depict complex systems where nodes correspond to entities and links encode interdependencies between them.  Generally, dynamics in networks is introduced via two different approaches. In the first approach, the links are assumed to be static, while the nodes are endowed with internal dynamics (epidemic spreading, opinion formation, signaling, synchronizing and so on). And in the second approach, nodes are treated as passive elements, and the main focus is on the evolution of network topology. 

More recently, it has been suggested that separating individual and network dynamics fails to capture realistic behavior of networks. Indeed, in most real--world networks  both  the attributes of individuals (nodes) and the topology of the network (links) evolve in tandem. Models of such adaptive co-evolving networks have attracted significant interest in recent years both in statistical physics~\cite{Pacheco2006,Gross2008,Castellano2009,Perc2010,Zschaler2012} and game theory and behavioral economics communities~\cite{Lazer2001,Jackson2002,Demange2005,Goyal2005,Goyal2009,Staudigl2013}.

To describe coupled  dynamics of individual attributes and network topology, here we suggest a simple model of a coevolving network that is based on the notion of interacting adaptive agents. Specifically,  we propose network--augmented multiagent systems where the agents play repeated games with their neighbors, and  adapt both their behaviors and the network ties depending on the outcome of their  interactions. To adapt,  the agents use a simple learning mechanism to reinforce (penalize) behaviors and  network links that produce favorable (unfavorable) outcomes. Furthermore, the agents use an action selection mechanism that allows one to control exploration/exploitation tradeoff via a temperature-like parameter.     

 We have previously demonstrated~\cite{Kianercy2012AAMAS} that the collective evolution of such a system can be described by appropriately defined replicator dynamics equations. Originally suggested in the context of evolutionary game theory (e.g., see Refs.~\cite{Hofbauer1998,Hofbauer2003}), replicator equations have been used  to model collective learning in systems of interacting self--interested agents~\cite{Sato2003}. Refrence~\cite{Kianercy2012AAMAS} provides a generalization to the scenario where the agents adapt not only their strategies (probability of selecting a certain action) but also their network structure (the set of other agents that play against).  This generalization results  in a system of coupled non-linear equations that describe the simultaneous evolution of agent strategies and network topology.   

Here we use the framework suggested in Ref.~\cite{Kianercy2012AAMAS} to examine the learning outcomes in networked games. We provide a comprehensive analysis of  three-player two-action games, which are the simplest systems  that exhibit non-trivial structural dynamics. We analytically characterize the rest-points and  their stability properties in the absence of exploration. Our results indicate that in the absence of exploration, the agents always play pure strategies even when the game allows mixed NE. For the general $n$-player case,  we find that the stable outcomes correspond to star-like motifs, and demonstrate analytically the stability of a star motif. We also demonstrate the instability of the symmetric network configuration where all the pairs are connected to each other with uniform weights. 

We also study the the impact of exploration on the co-evolutionary dynamics. In particular,  our results indicate that  there is a critical exploration rate above which the uniformly connected network is a globally stable outcome of the learning dynamics. 

The rest of the paper is organized as follows: we next derive the replicator equations characterizing the coevolution of the network structure and the strategies of the agents. In Sec.~\ref{sec:noexploration} we focus on learning without exploration, describe the NE of the game, and characterize the restpoints of learning dynamics according to their stability properties. We consider the  the impact of exploration on learning in Sec.~\ref{sec:explor} and provide some concluding remarks in Sec.~\ref{discussion}.

\section{Co-Evolving Networks via Reinforcement Learning } 
\label{sec:RL}

Let us consider a set of agents that play repeated games with each other. We differentiate  agents by indices $x, y, z, \ldots$. At each round of  the game, an agent has to choose another agent  to play with, and an action from the pool of  available actions. Thus,  time--dependent mixed strategies of agents   are characterized by a joint probability distribution over the choice of the neighbors and the actions.

We assume that the agents adapt to their environment through a simple reinforcement mechanism. Among different reinforcement schemes, here we focus on (stateless) $Q$-learning~\cite{Watkins1992}. Within this scheme, the   strategies of the agents are parameterized through, so-called $Q$ functions that characterize the relative utility of a particular strategy. After each round of game, the $Q$ functions are updated according to the following rule,  
\BEQ
\label{ararat}
Q_{xy}^i(t+1) = Q_{xy}^i(t) + \a [R_{x,y}^i(t) -  Q_{xy}^i(t)  ] 
\EEQ
where $R_{x,y}^i$($Q_{x,y}^i$) is the expected reward ($Q$ value) of agent $x$ for playing action $i$ with  agent $y$, and $\a$ is a parameter that determines the learning rate (which can be set to $\a=1$ without a loss of generality). 

Next, we have to specify how agents choose a neighbor and an action based on their $Q$ function.  Here we use the Boltzmann exploration mechanism  where the probability of a particular choice is given as~\cite{Sutton2000}
\BEQ
\label{dynamo}
p_{xy}^i = \frac{e^{\b Q_{xy}^i}}{\sum_{\yt,j}e^{\b Q_{x\yt}^j}}
\EEQ
where $p_{xy}^i$ is the probability that  agent $x$ will  play with agent $y$ and choose action $i$. 
Here  the inverse {\em temperature} $\b \equiv 1/T>0$ controls the tradeoff between exploration and exploitation; for $T\rightarrow  0$ the agents always choose the action corresponding to the maximum $Q$ value, while for $T\rightarrow \infty$ the agents' choices are completely random. 

We now assume that the agents interact with each other many times between two consecutive updates of their strategies. In this case, the reward of the $i$ th agent in Eq.~( \ref{ararat}) should be understood in terms of the {\em average reward}, where the average is taken  over the strategies of other agents, $R_{x,y}^i = \sum_{j} A^{ij}_{xy} p_{y x}^j$, where $A^{ij}_{xy}$ is the reward (payoff) of agent $x$ playing strategy $i$ against  agent $y$ who plays strategy $j$. Note that, generally speaking, the payoff  might be asymmetric.

We are interested in the continuous approximation to the learning dynamics.  Thus, we replace $t+1 \rightarrow t+\delta t$, $\alpha \rightarrow \alpha \delta t$, and take the limit $\delta t \rightarrow 0$ in Eq. (\ref{ararat}) to obtain 
\BEQ
\label{ararat2}
\dot{Q}_{xy}^i = \a [R_{x,y}^i -  Q_{xy}^i(t)  ] 
\EEQ
Differentiating Eq. (\ref{dynamo}), using Eqs.~(\ref{dynamo}) and (\ref{ararat2}), and scaling the time  $t\rightarrow \a\b t$,  we obtain the following replicator equation~\cite{Sato2003}:
\BEQ
\label{real}
\frac{ \dot{p}_{xy}^i}{p_{xy}^i} =    \sum_{j} A^{ij}_{xy} p_{y x}^j   -  \sum_{i,j,\yt} A^{ij}_{x\yt} p_{x \yt}^i  p_{\yt x}^j  + T  \sum_{\yt,j} p_{x\yt}^j \ln \frac{p_{x\yt}^j}{p_{xy}^i}
\EEQ
Equations~\ref{real} describe the collective adaptation of the Q--learning agents through repeated game--dynamical interactions. The first two terms indicate  that the probability of playing a particular pure strategy increases with a rate proportional to the overall goodness of that strategy, which mimics fitness-based selection mechanisms in population biology~\cite{Hofbauer1998}. The second term, which has an entropic meaning,   does not have a direct analog in population biology~\cite{Sato2003}. This term is due to the Boltzmann selection mechanism that describes the agents' tendency to randomize over their strategies. Note that for $T=0$  this term disappears, so the equations reduce to the conventional replicator system~\cite{Hofbauer1998}. 

So far, we have discussed learning dynamics over a general strategy space. We now make the assumption that the agents' strategies factorize as follows,
\BEQ
\label{barca}
p_{xy}^i = c_{xy} p_{x}^i \ , \ \ssum_{y} c_{xy} = 1, \  \ \ssum_{i} p_{x}^i = 1 .
\EEQ
Here $c_{xy}$ is the probability that the agent $x$ will initiate a  game with the agent $y$, whereas $p_x^i$ is the probability that he will choose action $i$.  Thus, the assumption behind this factorization is that  the probability that the agent will perform action $i$ is independent of  whom the game is played against. Substituting Eq.~(\ref{barca}) in Eq.~(\ref{real}) yields  
\BEA
\label{valencia}
\dot{c}_{xy} p_{x}^i + c_{xy} \dot{p}_{x}^i =   c_{xy}p_x^i  \biggl [ \sum_{j} a^{ij}_{xy} c_{yx}p_y^j  -  \sum_{i,y,j} a^{ij}_{x,y} c_{xy}c_{yx}p_x^i p_y^j  \nonumber \\ - T \biggl [  \ln c_{xy} + \ln p_{x}^i    - \sum_{y} c_{xy}\ln c_{xy}  - \sum_{j} p_x^j \ln p_{x}^j  \biggr ]  \biggr ]
\EEA
Next, we  sum both sides in Eq.~(\ref{valencia}), once over $y$ and then over $i$, and make use of the normalization conditions in Eq.~(\ref{barca})  to obtain the following coevolutionary dynamics of  action and connection probabilities:
\BEA
\frac{\dot{p}_{x}^i}{p_x^i} &=&  \sum_{\yt,j} A^{ij}_{x\yt} c_{x\yt}c_{\yt x}p_{\yt}^j  -  \sum_{i,j,\yt} A^{ij}_{x\yt} c_{x\yt}c_{\yt x}p_x^i p_{\yt}^j   \nonumber \\
&+&T\sum_j p_x^j \ln( p_x^j/p_x^i )
\label{sevilia1}\\
  \frac{\dot{c}_{xy}}{c_{xy}}  &=&  c_{yx}\sum_{i,j} A^{ij}_{xy} p_x^i p_y^j  - \sum_{i,j,\yt} A^{ij}_{x\yt} c_{x\yt}c_{\yt x}p_x^i p_{\yt}^j   \nonumber \\ 
&+&  T\sum_{\yt} c_{x \yt} \ln( c_{x \yt}/c_{xy} )\label{sevilia2}
\EEA 
 Equations~(\ref{sevilia1}) and ~(\ref{sevilia2}) are the replicator equations that describe the collective evolution of both the agents' strategies and the network structure.   

 The following remark is due: Generally, the replicator dynamics  in matrix games are invariant with respect to adding any column vector to the payoff matrix. However, this invariance does not hold in the present networked game. The reason for this is the following: if an agent does not have any incoming links (i.e., no other agent plays with him or her), then he always gets a zero reward. Thus, the zero  reward of an isolated agent serves  as a reference point. This poses a certain problem. For instance, consider a trivial game with a constant reward matrix $a_{ij}= P$. If $P>0$, then  the agents will tend to play with each other, whereas for $P<0$ they will try to avoid the game by isolating themselves (i.e.,  linking to agents that do not reciprocate).

To address this issue, we introduce an {\em isolation  payoff}  $C_I$ that an isolated agent receives at each round of the game.  It can be shown that the introduction of t his payoff merely subtracts $C_I$ from the reward matrix  in the replicator learning dynamics.\comment{Indeed, let us consider the expected payoff $r_{xy}^i$of  agent $x$ who chooses playmate $y$ and an action $i$. The probability that the agent $y$ plays with $x$ and chooses action $j$ is $p_{yx}^j$, so that the complementary probability of $y$ not playing with $x$ is $1-\ssum_{j}p_{yx}^j$. Thus, we have 
\BEA
r_{xy}^i&=&\ssum_{j}a_{ij}p_{yx}^j + C_{I}(1-\ssum_{j}p_{yx}) \nonumber \\
&=&\ssum_{j}(a_{ij}-C_{I})p_{yx}^j + C_I
\EEA 
}
Thus, we paramet rize  the game matrix as follows:
 \BEQ
 a_{ij}=b_{ij}+C_I
 \label{eq:cp}
 \EEQ
where matrix $B$ defines a specific game. 

Although it is beyond the scope of the present paper, an  interesting question is what the reasonable values for the parameter $C_I$ are. In fact, what is important is the value of $C_I$ relative to the reward at the corresponding Nash equilibria, i.e., whether {\em not playing at all} is better than {\em playing and receiving a potentially negative reward}. Different values of $C_I$ describe different situations. In particular, one can argue that certain social interactions are apparently characterized by large $C_I$, where not participating in a game is seen as a worse outcome  than participating and getting negative rewards. In the following, we  treat $C_I$ as an additional parameter that changes in a certain range,  and  examine its impact on the learning dynamics.

 \subsection{Two-action games}
  We focus on symmetric games where the reward matrix is the same for all pairs $(x,y)$, $A_{xy} =A$:
\BEQ
A= \left (   \begin{array}{ccc}
a_{11} & a_{12}\\
a_{21} & a_{22}
\end{array}
\right )
\EEQ
Let $p_{\alpha}$, $\alpha \in \{x,y,\dots, \}$, denote the probability for agent $\alpha$ to play action $1$ and $c_{xy}$ is the probability that  agent $x$ will initiate a  game with the agent $y$. For two action games, the learning dynamics Eqs.~(\ref{sevilia1}) , and ~(\ref{sevilia2}) becomes:
\BEA
\label{TwoactionREP1}
\frac{ \dot{p}_{x}}{p_{x}(1-p_{x}) } &=&  \sum_{\yt}(ap_{\yt}+b) c_{x\yt}c_{\yt x}+ T \log\frac{1-p_x}{p_x}\\
\label{TwoactionREP2} \frac{ \dot{c}_{xy}}{c_{xy}} &=&   r_{xy}- R_{x} +T \sum_{\yt} c_{x \yt} \ln \frac{c_{x \yt}}{c_{xy}}  
\EEA
where
\BEA
\label{TwoactionR}
r_{xy}&=&c_{yx}(ap_{x}p_{y}+b p_{x}+d p_{y}+a_{22}) ~~~\\ 
R_{x}&=&\sum_{\yt}(ap_{x}p_{\yt}+bp_{x}+dp_{\yt}+a_{22})c_{x\yt}c_{\yt x}
\EEA
Here we have defined the  following parameters:
\BEA
 a&=&a_{11} - a_{21} - a_{12} + a_{22} \label{cof1} \\
b&=&a_{12} - a_{22} \label{cof2}  \\
d&=&a_{21}-a_{22} \label{cof3}
\EEA

\begin{figure}[!t]
\centering
\includegraphics[width=0.4\textwidth]{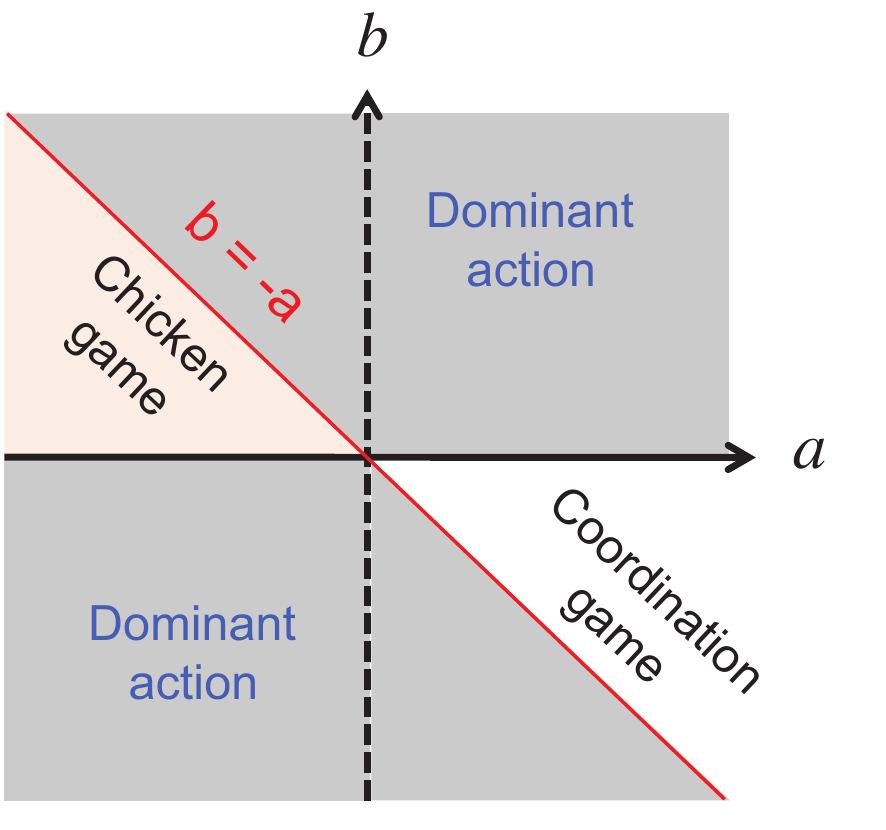}
\caption{(Color online)  Categorization of two-action games based on the reward matrix structure in the $(a,b)$ plane. }  
\label{gamecat}
\end {figure}

The parameters $a$ and $b$ allow a categorization of two action games as follows (Fig. \ref{gamecat}): 
\begin{itemize}
\item \it{dominant action games}: $ -\frac{b}{a}>1\hspace{0.2cm}or \hspace{0.25cm}-\frac{b}{a}<0 $
\item \it{coordination game}: $ a>0 , b<0 \hspace{0.25cm} and \hspace{0.25cm}1\ge -\frac{b}{a} $
\item \it{anti-coordination (Chicken) game}: $ a<0 , b>0 \hspace{0.25cm} and \hspace{0.25cm}1\ge -\frac{b}{a} $
\end{itemize}

 Before proceeding further, we elaborate on the connection between the rest points of the replicator system for $T=0$ and the game-theoretic notion of NE (NE)~\footnote{Recall that a joint strategy profile is called NE  if no agent can increase his expected reward by {\em unilaterally}  deviating from the equilibrium.}. For  $T=0$ (no exploration) in the conventional replicator equations, all NE are necessarily the rest points of the learning dynamic. The inverse is not true - not all rest points correspond to NE - and only the stable ones do.  Note  that in the present model the first statement does not necessarily  hold. This is because we have assumed the strategy  Eq.(~\ref{barca}), due to which equilibria where the agents adopt different strategies with different players are not allowed. Thus, any NE that do not have the factorized form simply cannot be described in this framework. The second statement, however, remains true, and stable rest points do correspond to NE.  

  \section{ Learning without exploration }
  \label{sec:noexploration}
For $T=0$, the learning dynamics  Equations ~(\ref{TwoactionREP1}),  (\ref{TwoactionREP2}) attain the following form:
\BEA
\frac{ \dot{p}_{x}}{p_{x}(1-p_{x})}& =& \sum_{\yt}(ap_{\yt}+b) c_{x\yt}c_{\yt x}\label{REP1} \\
\frac{ \dot{c}_{xy}}{c_{xy}} &=&   r_{xy}- R_{x}\label{REP2} 
\EEA

Consider the dynamics of the strategies given by Eq.~\ref{REP1}. Clearly, the vertices of the simplex, $p_x=\{0,1\}$ are the rest points of the dynamics. Furthermore, in case the game allows a mixed NE,  then the configuration where all the agents play the mixed NE $p_x=-b/a$ is also a rest point of the dynamics. As is shown below, however, this configuration is not stable, and for $T=0$, the only stable configurations correspond to the agents playing pure strategies. 

\subsection{Three-player games}
\label{three player}
We now consider the case of three players in two-action games. This scenario is simple enough for studying it comprehensively, yet it still has non-trivial structural dynamics, as we demonstrate below. 

 \begin{figure}[!t]
\centering
    \includegraphics[width = 0.45\textwidth]{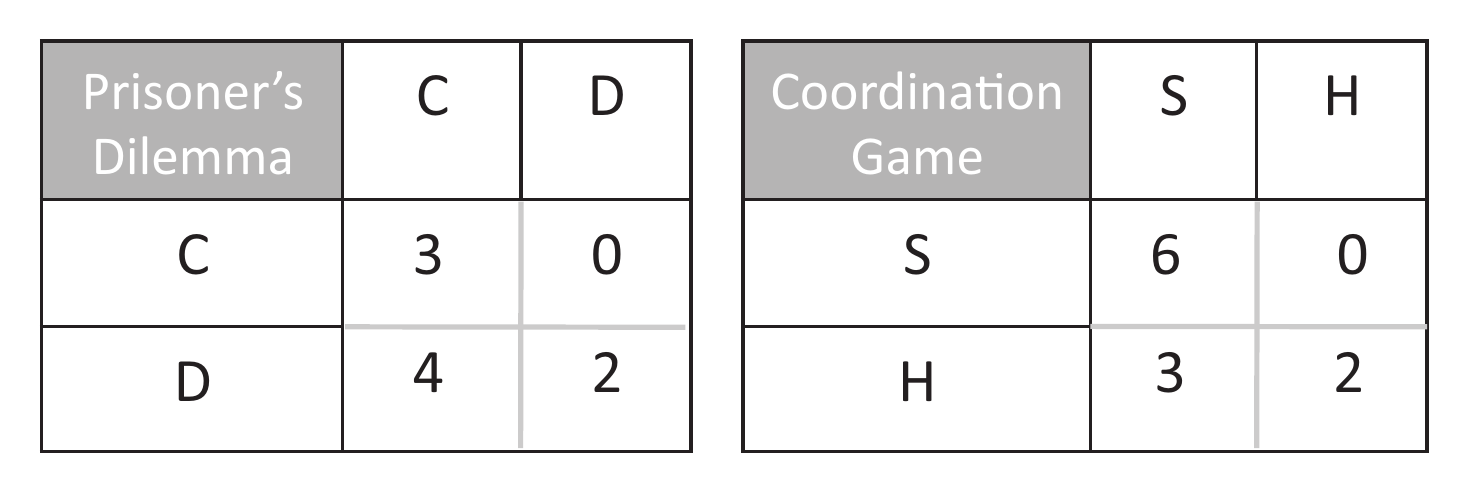}
       \caption{Examples of reward matrices for typical two-action games.}
        \label{fig7}
\end{figure}

 \subsubsection{ Nash equilibria}  
 \label{sec:ne}

\begin{figure}[!]
\centering
    \includegraphics[width=0.23\textwidth]{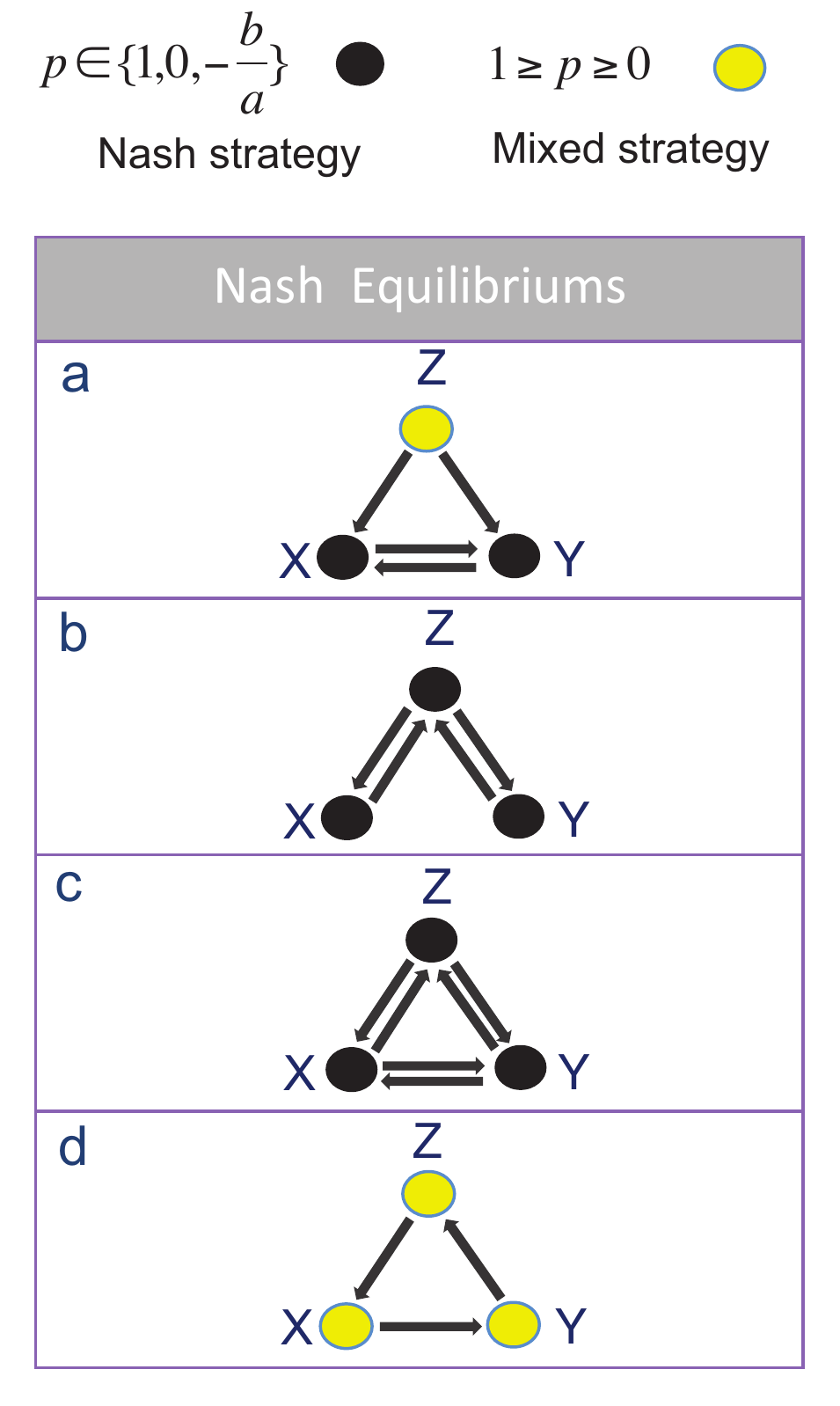}
          \caption{ (Color online) Three-player network NE  for prisoner's dilemma and the coordination game; see the text for more details. 
          }
\label{fig:nash}
\end{figure}

We start by examining  the NE for two classes of two-action games, prisoner dilemma (PD) and a coordination game.~\footnote{The behavior of the coordination and anti-coordination games are qualitatively similar in the context of the present work, so here we do not consider the latter.} In PD, the players have to choose between {\em Cooperation} and {\em Defection}, and the payoff matrix  elements satisfy  $b_{21}>b_{11}>b_{22}>b_{12}$;  (see Fig.~\ref{fig7}). In a two-player PD game, defection is a {\em dominant} strategy -- it always  yields a better reward regardless of the other player choice -- thus, the only NE is a mutual defection.  And in coordination game, the  players have an incentive to select the same action. This game has two pure NE, where the agents choose the same action,  as well as  a mixed NE. In a general coordination game the reward elements have the  relationship $b_{11}>b_{21}  , b_{22}>b_{12}$ (see Fig.~\ref{fig7}).

In the three-agent scenario, a simple analysis yields four possible network topologies corresponding to NE depicted in   Fig.~\ref{fig:nash}. In all of those configurations, the agents that are not isolated select strategies that correspond to two-agent NE. Thus, in the case of PD, non-isolated agents always defect, whereas for the coordination game, they can select one of three possible NE.  We now examine those configurations in more details. 

\noindent {\bf Configuration I} In this configuration, the agents $x$ and $y$ play only with each other, whereas agent is $z$ s isolated: $c_{xy}=c_{yx}=1$. Note that for this to be a NE, agents $x$ and $y$ should not be ``tempted" to switch and play with the agent $z$. For instance, in the case of PD, this yields $p_z b_{21}<b_{22}$, otherwise players $x$ and $y$ will be better of linking with the isolated agent $z$  and exploiting his cooperative behavior.~\footnote{Note that  the dynamics will eventually lead to a different rest point where $z$ is now plays defect with both $x$ and $y$.} \\
 
 \noindent  {\bf Configuration II}  In the second configuration, there is  a central agent ($z$) who plays with the other two: $c_{xz}=c_{yz}=1 , c_{zx}+c_{zy}=1$. Note that this configuration is continuously degenerate as the central agent can distribute his link weight arbitrarily among the two players. Additionally, the isolation payoff should be smaller then than the reward at the equilibrium (e.g., $b_{22}>C_{I}$ for PD). Indeed, if the latter condition is reversed, then one of the agents, say $x$, is better off linking with $y$ instead of $z$, thus ``avoiding" the game altogether.  \\
 
\noindent  {\bf Configuration III:} The third configuration corresponds to a uniformly connected networks where all the links have the same weight $c_{xy}=c_{yz}=c_{cx}=\frac{1}{2}$. It is easy to see that when all three agents play NE strategies, there is no incentive to deviate from the uniform network structure.  \\

\noindent  {\bf Configuration IV:} Finally, in the last configuration none of the links are reciprocated so that the players do not play with each other: $c_{xy}c_{yx}=c_{xz}c_{zx}=c_{yz}c_{zy}=0$. This cyclic network is a NE when the isolation payoff  $C_{I}$ is greater than the expected reward of playing NE in the respective game.

\subsubsection{Stable rest points of learning dynamics} 
\label{Stability}
The factorized NE  discussed in the previous section are the rest points of the replicator dynamics. However, not all of those rest points are stable, so that not all the equilibria can be achieved via learning.  We now discuss the stability property of the rest points.

One of the main outcomes of our stability analysis is that at $T=0$ the symmetric network configuration is not stable. This is in fact a more general results that applies to $n$-agent networks, as is shown in the next  section. As we will demonstrate later, the symmetric network can be stabilized when one allows exploration.

The second important observation  is that even when the game allows mixed NE, such as in the coordination game, any network configuration where the agents play mixed strategy is unstable for $T=0$ (see Appendix~\ref{app:stability}). Thus, the only outcome of the  learning is a configuration where the agents play pure strategies. 

\begin{figure}[!t]
\centering
 \subfigure{
    \includegraphics[width=0.48\textwidth]{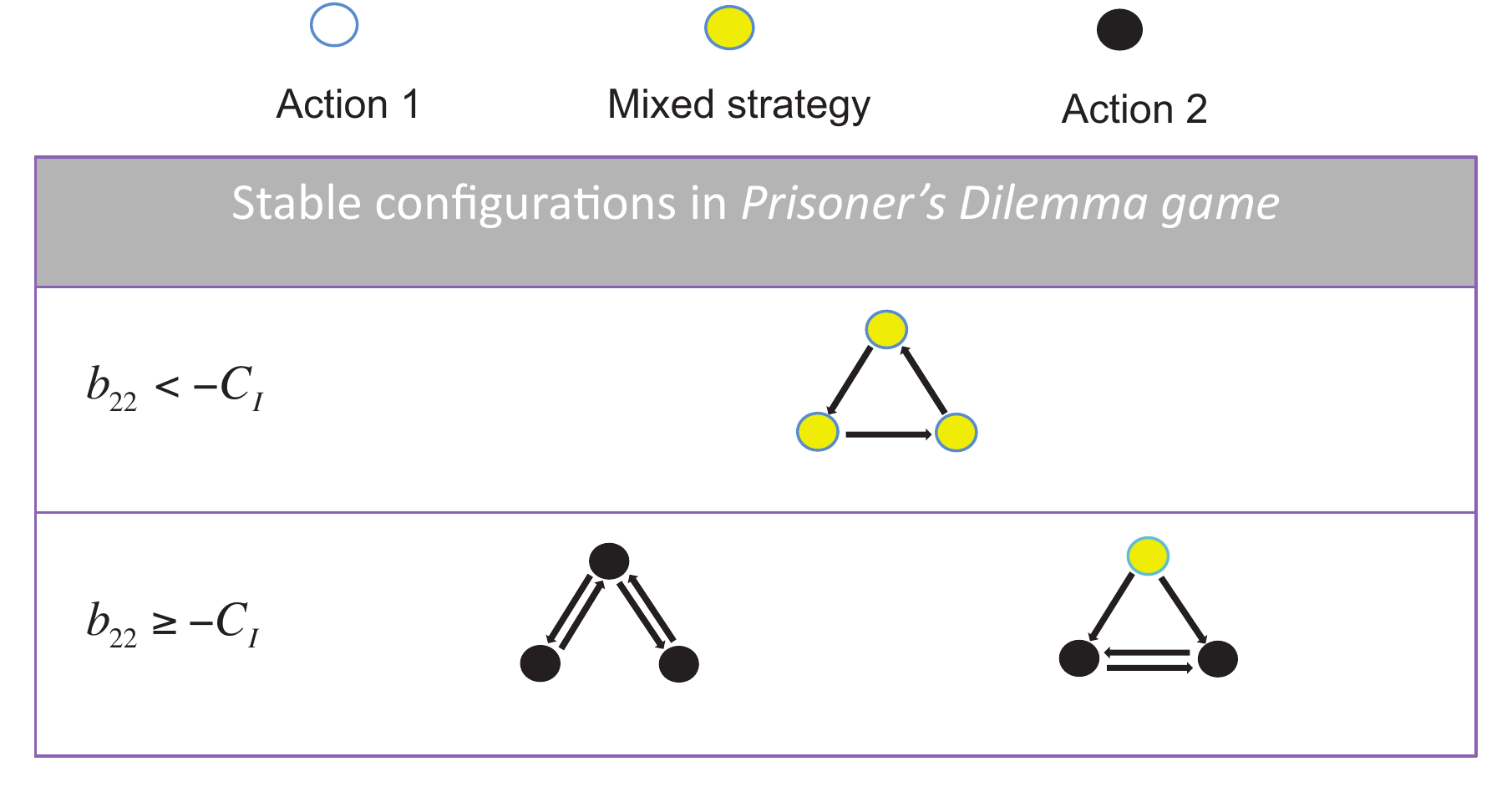}\label{fig3a}
    }   \\
       \subfigure{
   \includegraphics[width=0.48\textwidth]{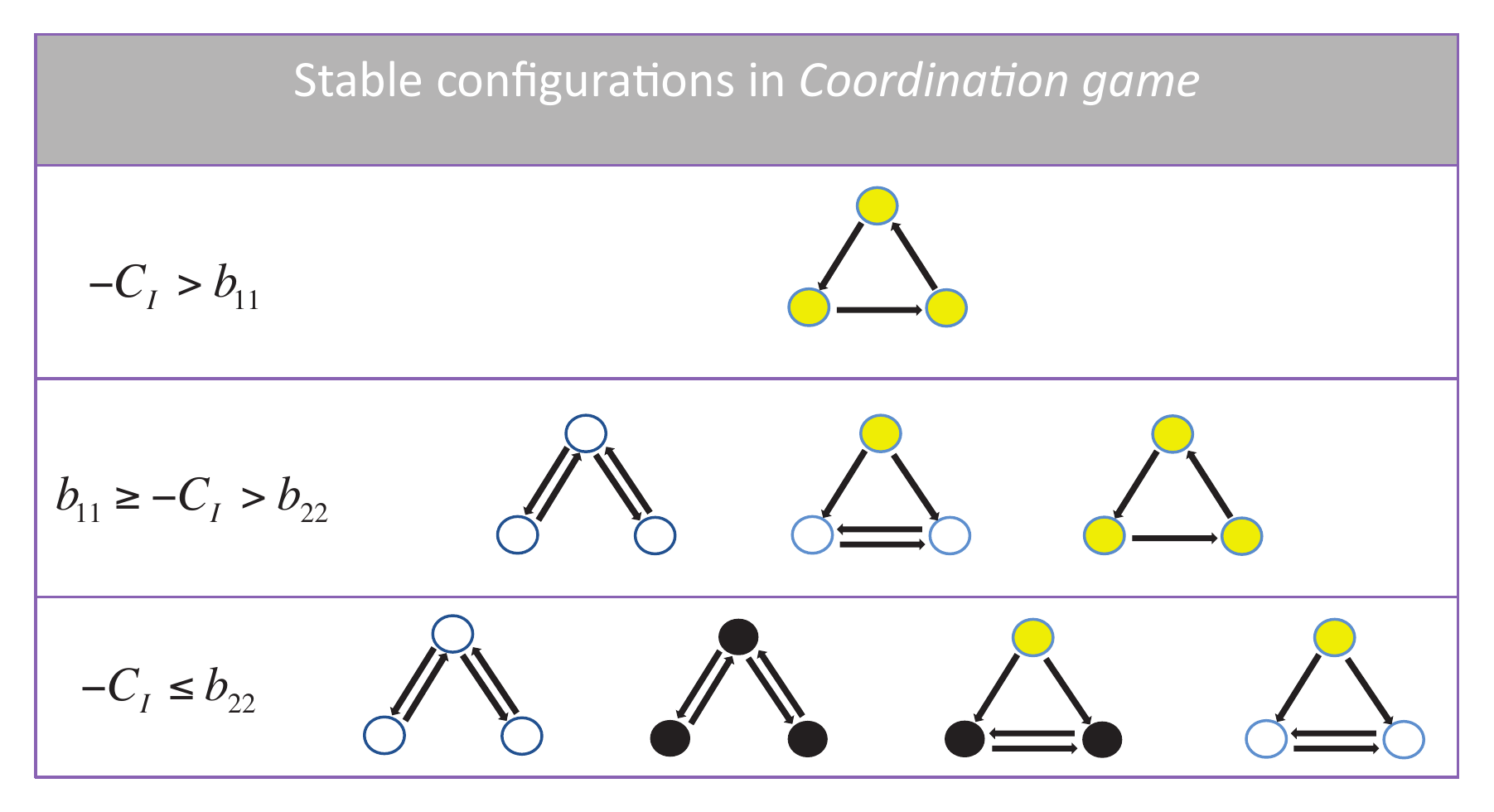}\label{fig3b}
    }      
\caption{ (Color online)  Stable rest points of the learning dynamics for PD (upper panel)  and the coordination game (lower panel). 
}
\label{fig3}
\end{figure}

The surviving (stable) configurations are listed in Fig.~\ref{fig3}. Their stability can be established by analyzing the eigenvalues of the corresponding Jacobian. Consider, for instance, the configuration with one isolated player. The corresponding eigenvalues are 
\begin{eqnarray*}
\begin{aligned}
\lambda_{1}&=r_{xz}-r_{xy}~,~
\lambda_{2}=r_{yz}-r_{yx}~,~
\lambda_{3}=0\\
\lambda_{4}&=(1-2p_{x})(r_{x}^{1}-r_{x}^{2})<0~,~\\
\lambda_{5}&=(1-2p_{y})(r_{y}^{1}-r_{y}^{2})<0 ~,~
\lambda_{6}=0
\end{aligned}
\end{eqnarray*}
For PD this configuration is marginally stable when agents $x$ and $y$ play defect and $r_{xy}>0$ and $r_{yx}>0$.  It happens only when $b_{22}\ge -C_{I}$ which means that the isolation payoff should be less than the expected reward for defection. Furthermore, one should also have $ r_{xz}<r_{xy}~ ,~ r_{yz}< r_{yx}$, which indicates that the neither $x$ nor $y$ would get a better expected reward by switching and playing with $z$ (e.g., condition for NE). And for the coordination  game , assuming that $b_{11}>b_{22}$  this configuration is stable when $b_{11}\ge -C_{I}>b_{22} ~ and ~b_{22}\ge -C_{I}.$

Similar reasoning can be used for the other configurations shown in Fig.~\ref{fig3}. Note, finally, that there is a coexistence of multiple equlibria for  range of parameter, except when the isolation payoff is sufficiently large, for which the cyclic (non-reciprocated) network is the only stable configuration.

\subsection{ $n$-player games} 
\label{n player}
\begin{figure}[t]
\centering
\includegraphics[width=0.45\textwidth]{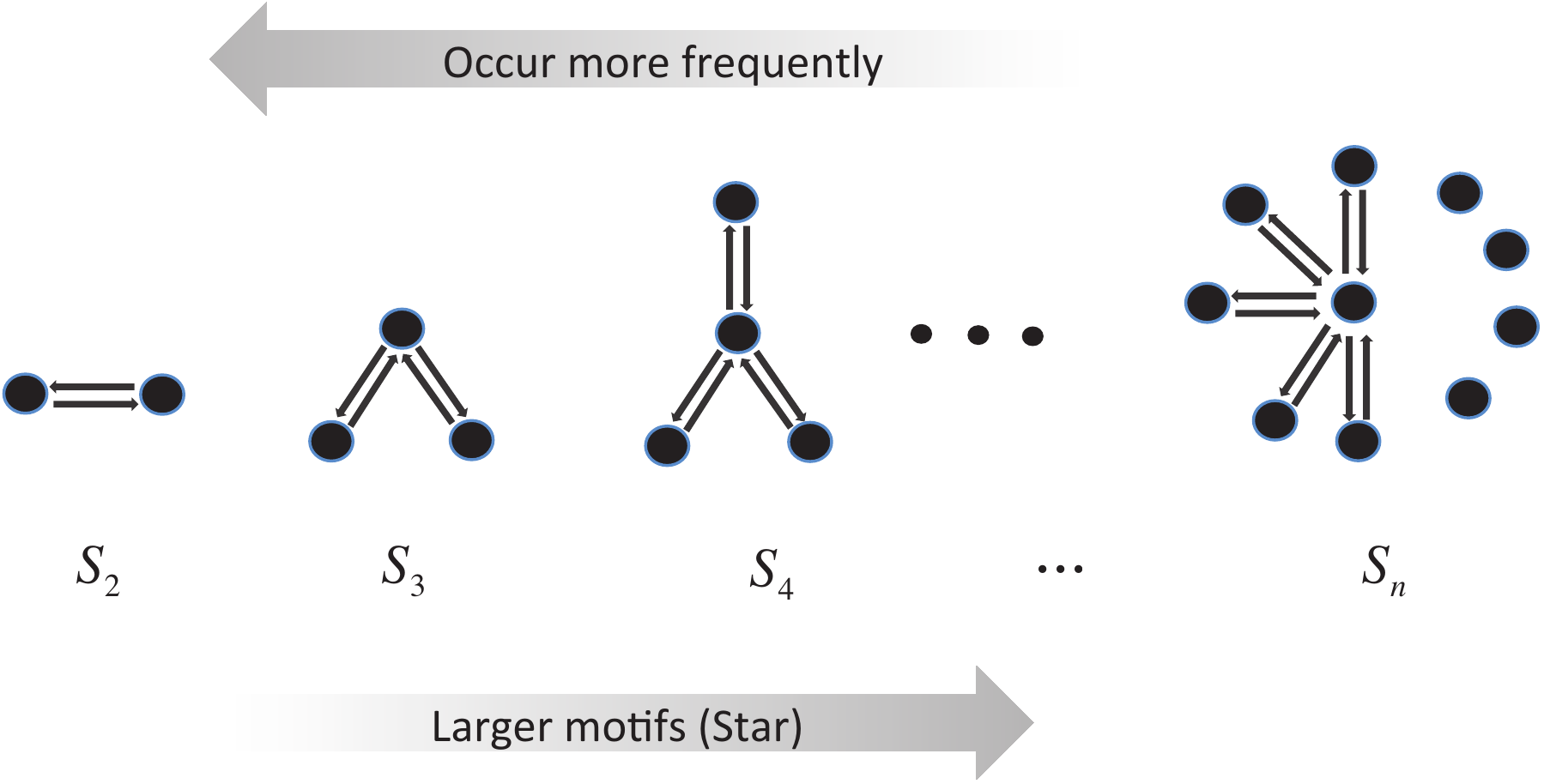}
\caption{Observed stable configurations of co-evolutionary dynamics for $T=0$.  }  
\label{fig4}
\end {figure}

In addition to the three agent scenario, we also examined the  co-evolutionary dynamics of general $n$-agent systems,  using both simulations and analytical methods. We observed in our simulations that  the stable outcomes of the learning dynamic consist of $star$ motifs $S_{n}$ (Fig.~\ref{fig4}), where a central node of degree $n-1$ connects to $n-1$ nodes of degree $1$.~\footnote{This is true when the isolation payoff is smaller compared to the NE payoff. In the opposite case the dynamics settles into a configuration without reciprocated links.}  Furthermore,  we observed that the  basin of attraction of motifs shrinks as   motif size  grows, so that smaller motifs are more frequent.


We now demonstrate  the stability of  the star motif $S_{n}$  in  $n$ player two action games. Let player  $x$ be the central player, so that  all other players are only connected to  $x$, $c_{\alpha x}=1$. Recall that the Jacobian of the system is a block diagonal matrix with blocks $J_{11}$ with elements $\frac{\partial \dot{c}_{ij}}{\partial c_{mn}}$  and $ J_{22}$ with has elements as $\frac{\partial \dot{p}_{m}}{\partial p_{n}}$( see Appendix~\ref{app:stability}). When all players play a pure strategy $p_{i}=0,1$ in a star shape motif, it can be 
shown that  $J_{22}$  is diagonal matrix with diagonal elements of form  $(1- 2 p_{x})\sum_{\tilde{y}} (a p_{\tilde{y}}+b) c_{x\tilde{y}} c_{\tilde{y}x}$, whereas 
$J_{11}$ is  an upper triangular matrix, and its  diagonal elements  are  either zero or have the form $- (a p_{x} p_{y} + b p_{x} +d p_{y} +a_{22}) c_{xy}$ where $x$ is the central player.

For the Prisoner's Dilemma, the  Nash Equilibrium corresponds to choosing the second  action (defection) , i.e. $p_{\alpha}=0$. Then the diagonal elements of $J_{22}$, and thus its eigenvalues,  equal $b c_{x\tilde{y}}$.  $J_{11}$, on the other hand, has $n^2-2n$ eigenvalues , $(n-1)$ of them are  zero and the rest have the form of $ \lambda=-a_{22}c_{x \tilde{y}}$. Since for the Prisoner's Dilemma one has $b<0$ then the start structure is stable as long as  $b_{22}>C_I$. 

A similar reasoning can be used for the  Coordination game, for which one has $b<0$ and  $a+b>0$. In this case, the star structure is stable  when either  $b_{11}>-C_{I}$ or $b_{22}>-C_{I}$, depending on whether  the agents coordinate on the first or second actions, respectively.

 We conclude this section by elaborating on the (in)stability of the $n$-agent symmetric network configuration, where each agent is connected to all the other agents with the same connectivity $\frac{1}{n-1}$. As shown in Appendix~\ref{app:Sym}, this configuration can be a rest point of the learning dynamics Eq.~(\ref{REP1})  only when  all agents play the same strategy, which is  either $0, 1$ or  $-b/a$. Consider now the first block of the Jacobian in Eq.~\ref{jacob}, i.e. $J_{11}$. It can be shown that the diagonal elements of $J_{11}$ are identically zero, so that $Tr (J_{11}) = 0$. Thus, either all the eigenvalues of $J_{11} $ are zero (in which case the configuration is marginally stable), or there is at least one eigenvalue that is positive, thus making the symmetric network configuration unstable at $T=0$.


\section{Learning with Exploration}
\label{sec:explor}
In this section we consider the replicator dynamics for non-vanishing exploration rate $T>0$. For two agent games, the effect of the exploration has been previously examined in Ref.~\cite{Kianercy2012}, where it was established that for a class of games with multiple Nash equilibria  the asymptotic behavior of learning dynamics  undergoes a drastic changes at critical  exploration rates and only one of those equilibria survives. Below, we study the impact of the exploration in the current networked version of the learning dynamics. 

For  3-player, 2- action games we have six independent variables $ p_{x}, p_{y}, p_{z}, c_{xy}, c_{yz}, and c_{zx}$. The strategy variables evolve  according to the following equations:
\BEA
\frac{\dot{p_x}}{(1-p_{x})p_{x}}&=&(ap_{y}+b) w_{xy}+(ap_{z}+b) w_{xz}+T \log \frac{1-p_{x}}{p_x} \nonumber  \\
\frac{\dot{p_y}}{(1-p_{y})p_{y}}&=&(ap_{z}+b) w_{yz}+(ap_{x}+b) w_{xy}+T \log \frac{1-p_{y}}{p_y}  \nonumber  \\
\frac{\dot{p_z}}{(1-p_{z})p_{z}}&=&(ap_{x}+b) w_{xz}+(ap_{y}+b) w_{yz}+T \log \frac{1-p_{z}}{p_z} \nonumber  \\
\frac{\dot{c}_{xy}}{c_{xy}(1-c_{xy})}&=&r_{xy} - r_{xz} +T \log \frac{1-c_{xy}}{c_{xy}} \nonumber  \\
\frac{\dot{c}_{yz}}{c_{yz}(1-c_{yz})}&=&r_{yz} - r_{yx} +T \log \frac{1-c_{yz}}{c_{yz}} \nonumber \\
\frac{\dot{c}_{zx}}{c_{zx}(1-c_{zx})}&=&r_{zx} - r_{zy} +T \log \frac{1-c_{zx}}{c_{zx}} \nonumber 
\EEA
Here we have defined  $w_{xy}=c_{xy}(1-c_{yz})$, $w_{xz}=(1-c_{xy})c_{zx}$, and $w_{yz}=c_{yz}(1-c_{zx})$, and $a, b, and d$ are defined in Eqs.~\ref{cof1},~\ref{cof2} and~\ref{cof3}. 

 \begin{figure}[t!]
\centering   
\subfigure[]{
    \includegraphics[width = 0.46\textwidth]{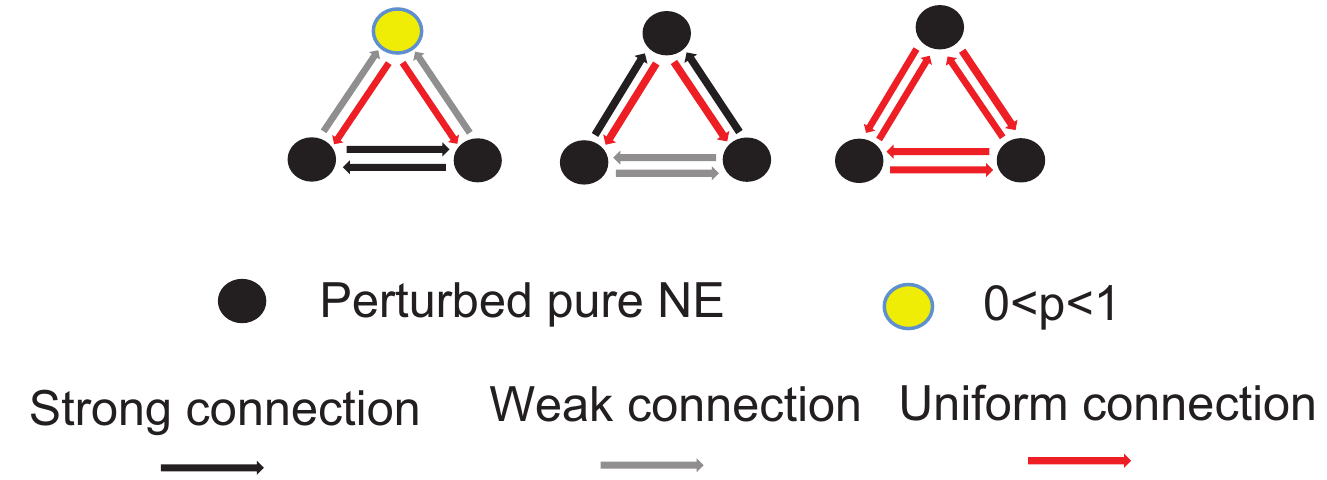} \label{restpoint}
    }
    \subfigure[]{
    \includegraphics[width = 0.435\textwidth]{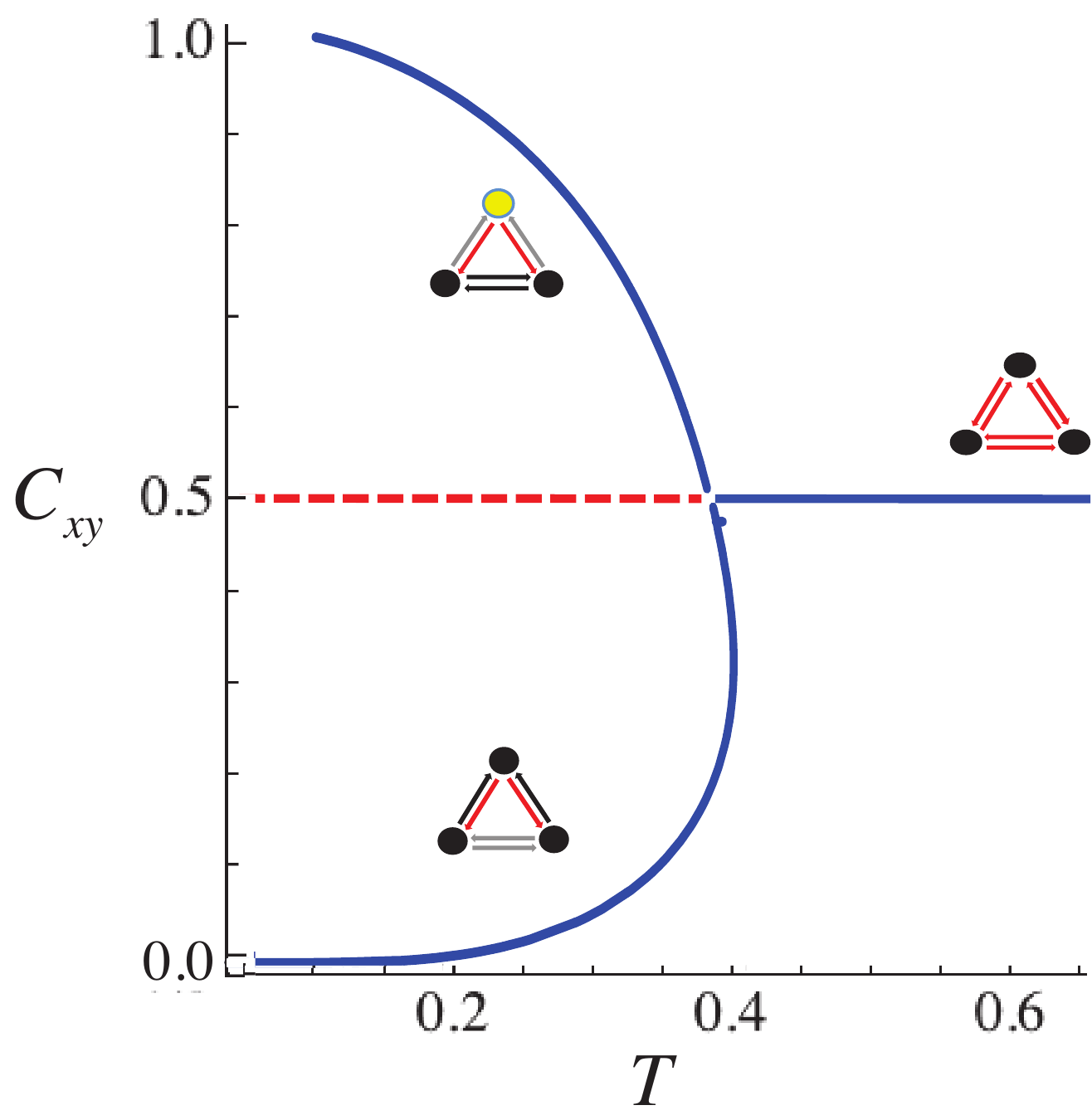} \label{connection_bifo}
    }  
   \caption{ a) (Color online)  Possible network configurations for three-player  PD (Fig.~\ref{fig7})., (b) Bifurcation diagram for varying temperature. Two  blue solid lines correspond to the configurations with one isolated agent and one central agent. The symmetric network configuration  is  unstable  at low temperature (red line),  and becomes globally stable above a  critical temperature.}
\end{figure}

 Figure ~\ref{restpoint}  shows  three possible network configurations that correspond to the fixed points of the above dynamics. The first two configurations are perturbed version of a star motif ( stable solution for $T=0$), whereas the third one corresponds to a symmetric network where all players connect  to the other players with equal link weights.  
 
 Furthermore, in Fig.~\ref{connection_bifo} we show the behavior of the learning outcomes for a PD game, as one varies the temperature. For sufficiently small $T$, the only stable configurations are the perturbed star motifs, and the symmetric network is unstable. However, there is a critical value $T_c$  above which  the symmetric network becomes globally stable. 
 
 \begin{figure}[!t]
\centering   
 \includegraphics[width = 0.4\textwidth]{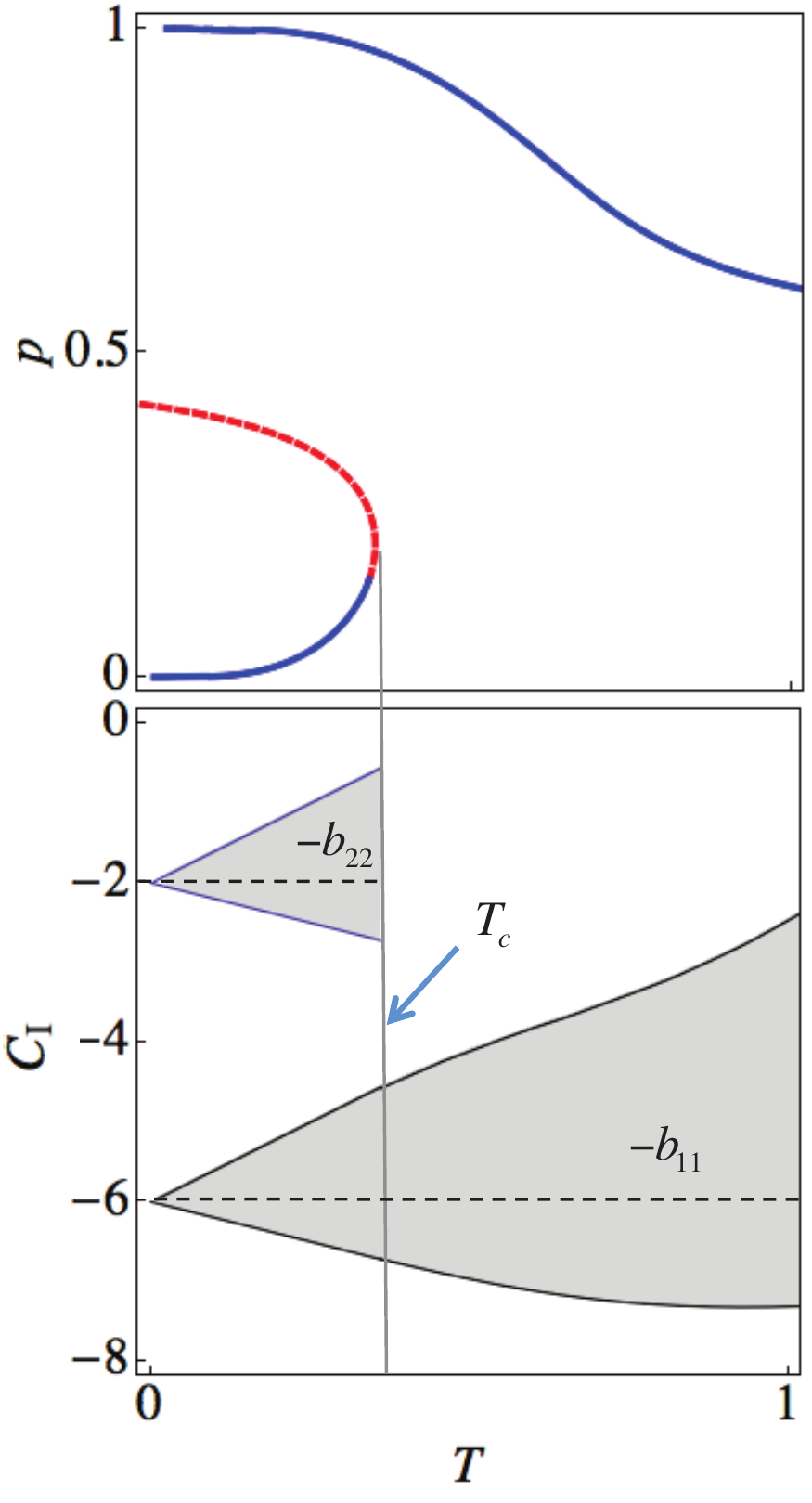} 
   \caption{(Color online)  Impact of the exploration on the stable outcomes of a  coordination game in Fig.~(\ref{fig7}). The top panel shows the bifurcation of strategy $p$ versus $T$, whereas the bottom panel shows the stability region of the symmetric network configuration in the $C_{I}-T$ plane. Here the critical temperature is $T_{c}=0.36$.}
   \label{fig9}
\end{figure}

Next, we consider the stability of the symmetric networks. As shown in Appendix~\ref{app:Sym}, the only possible solution in this configuration is when all the agents play the same strategy, which can be found from the following equation:
\BEQ
\label{REP5}
(ap+b) =2  T  \log \frac{p}{1-p}
\EEQ
The behavior of this equation (without the factor $2$ in the right-hand side)  was analyzed in details in Ref.~\cite{Kianercy2012}. In particular, for games with a single NE, this equation allows a single solution that corresponds to the perturbed NE. For games with multiple equilibria, on the other hand, there is a critical exploration rate $T_c$: For $T<T_c$ there are two stable solutions and one unstable solution, whereas for $T\ge T_c$ there is a single globally stable solution.

We use these insights to examine the stability of the symmetric network configuration for the coordination game, depending on the parameters $T$ and $C_I$; see Appendix~\ref{app:Dyn}. 
In this example $a=5$ , $b=-2$  and $ d=1$ for all three agents. Figure~\ref{fig9} shows the bifurcation diagram of  $p$ (probability of choosing the first action) plotted versus $T$.   Below the critical temperature, there are   three three solutions,  two of which (that correspond to the perturbed pure NE) are stable. And Fig.~(\ref{fig9})  shows the domain of $T$  and $C_I$ for  stable homogenous equilibrium. When $T \to 0$, the domain of $C_I$  shrinks  until it becomes a point at $T=0$ where $-C_I$ is  equal to the  NE reward (Fig.~\ref{fig9}).

\section{Discussion}
\label{discussion}

We have studied the co-evolutionary dynamics of strategies and link structure in a network of reinforcement-learning agents.  By assuming that the agents' strategies allow appropriate factorization, we derived a system of a coupled replicator equations  that describe the mutual evolution of agent behavior and network topology. We used these equations to fully characterize the stable learning outcomes in the case of three agents and two action games. We also established some analytical results for the more general case of $n$-player two-action games.

We demonstrated that in the absence of any strategy exploration (zero temperature limit) learning leads to a network composed of star-like motifs. Furthermore, the agents on those networks play only pure NE, even when the game allows a mixed NE.  Also, even though the learning dynamics allows  rest points with a uniform network (e.g., an agent  plays with all the other agents with the same probability) , those equilibria are not stable at $T=0$.  The situation changes when the agents  explore their strategy space. In this case, the stable network structures undergo bifurcation as one changes the exploration rate. In particular, there is a critical exploration rate above which the uniform network becomes a globally stable outcome of the learning dynamics. 

We note that the main premise behind the strategy factorization use here is that the agents use the same strategy profile irrespective of whom they play against.  While this assumption is perhaps valid  under certain circumstances, it certainly has its limitations that need to be studied further through analytical results and empirical data. Furthermore, the other extreme where the agent employs unique strategy profiles for each of his partners does not seem very realistic either, as it would impose considerable cognitive load on the agent. A more realistic assumption is that the agent has a few strategy profile that roughly corresponds to the type of agent he is interacting with. The approach presented here can be, in principle, generalized to the latter scenario.         

\section{ Acknowledgments}
We thank Armen Allahverdyan  for his comments and contributions during the initial phase of this work. This research was supported in part by the National Science Foundation under Grant No. 0916534 and the U.S. AFOSR MURI under Grant No. FA9550-10-1-0569.

\appendix

\section{Local Stability Analysis of the Rest Points}
\label{app:stability}
To study the local stability properties of the rest points in the system  given by Eqs.\ref{REP1} and ~\ref{REP2} , we need to analyze the eigenvalues of the corresponding Jacobian matrix. For $n$-player two-action game, we have  $n$ action variables and $l=n(n-2)$ link variables, so that the total number of independent dynamical variables is $n+l=n(n-1)$. We can represent the Jacobian as follows,
\BEQ
J= \left (   \begin{array}{cc}
\frac{\partial \dot{c}_{ij}}{\partial c_{mn}} & \frac{\partial \dot{c}_{ij}}{\partial p_{m}} \\ [4.5mm]
\frac{\partial \dot{p}_{m}}{\partial c_{ij}}&\frac{\partial \dot{p}_{m}}{\partial p_{n}} \\
\end{array}
\right )=
\begin{pmatrix}
J_{11} & J_{12}\\ 
 J_{21}& J_{22} 
\end{pmatrix}
\label{jacob}
\EEQ
Here the diagonal blocks $J_{11}$ and $J_{22}$ are  $l\times l$ and $n\times n$ square matrices, respectively. Similarly, $J_{12}$ and $J_{21}$ are $l\times n$ and $n\times l$ matrices, respectively. 

In the most general case, the full analysis of the  Jacobian is intractable. However, the problem can be simplified for $T=0$. Indeed, consider the lower off-diagonal block of the Jacobian, $J_{21}$, the elements of which have the form 
\BEQ
\frac{\partial \dot{p}_{i}}{\partial c_{ij}}=p_{i}(1-p_{i})c_{ji}(ap_{i}+b)
\EEQ
Consulting the rest point condition given by Eqs.~\ref{REP1}, one can see that  $J_{21}$ is identically zero. By using the {\em block matrix  determinant} identity, the characteristic polynomial of the Jacobian assumes the following factorized form
\begin{align}
p(\lambda)=\bf{det}(J_{11}-\lambda \bf{I})\bf{det}(J_{22}-\lambda \bf{I})=0
 \label{Char}
\end{align}
The above factorization facilitates the stability analysis for certain cases that we now focus on:

\paragraph{(In)Stability of mixed strategies for $T=0$ }
Let us show that the configurations where the agents mix either on their actions or links cannot be stable at $T=0$. Here  we just need to consider the submatrix  $J_{22}$. We now show that this matrix always has at least one positive eigenvalues when players adopt the mixed NE  $p=-b/a$. Indeed, it can be shown that $J_{22}$ is a non-zero matrix with zero diagonal elements. Recall that for any square matrix $A$ the $Tr(A)=\sum \lambda_{i}$    then  $Tr(J_{11})=0$  means at least one of its eigenvalues is  always positive,  so that the  mixed  Nash configuration is  unstable. The same line of reasoning can be applied to the configuration where the agents mix over the links.

\section{Agent Strategies in Symmetric Networks}
\label{app:Sym}
Let us  consider a two-action  $n$-players  game. Each player  $i$ chooses  action one  with probability $p_i$. Here we prove that  player $n$ and player $n-1$ in a homogenous network have the same strategy, i.e., $p_n=p_{n-1}$.  Consider  Eq.~(\ref{TwoactionREP1}) for players $n$, $ n-1$ and $n-2$,
\BEA
p_{1}+p_{2}+\cdots+p_{n-2}+p_{n-1}=k \log \frac{p_n}{1-p_{n}} -c
\label{sym1}
\EEA
\BEA
p_{1}+p_{2}+\cdots+p_{n-2}+p_{n}=k \log \frac{p_{n-1}}{1-p_{n-1}}-c
\label{sym2}
\EEA
where 
\BEA
K=-\frac{T (n-1)^2}{a} \ , \  c=\frac{b(n-1)}{a} 
\label{cofapx1}
\EEA
 Also, let us  define a function $g$ as
 \BEA
g(p_n)=x_{n}+k \log \frac{p_n}{(1-p_{n})}
\EEA

 Now , by  subtracting the two Eqs.(\ref{sym1}) and (\ref{sym2}), we have $g(p_n)=g(p_{n-1})$. Since  $0<p_{i}<1$ , then function $g$ is a monotonic  function, so $g(p_n)=g(p_{n-1}) \leftrightarrow  p{n}=p_{n-1}$. By repeating the same reasoning for the remaining  $p_i$ one can prove  that $ p_1=p_2=\cdots=p_n$.

\section{ Stability of Symmetric three-player network }
\label{app:Dyn}
For three-player  two-action games, the Jacobian corresponding to the symmetric network configuration consists of the following blocks:  
\BEA
       J_{11}&=&\begin{pmatrix}
                   -T&-v&-v \\
                   -v &-T &-v \\
                   -v & -v&-T
\end{pmatrix}, \\
       J_{12}&=&\begin{pmatrix}
                   0&m&-m\\
                   -m &0 &m \\
                   m& -m&0
\end{pmatrix},\\
       J_{21}&=&\begin{pmatrix}
                   0&-g&g \\
                   g&0&-g \\
                   -g&g&0
\end{pmatrix}, \\
       J_{22}&=&\begin{pmatrix}
                   -T&k&k \\
                   k&-T &k\\
                   k&k&-T 
\end{pmatrix},
\EEA
where we have defined
 \BEA
 v&=&\frac{ap^2+bp+dp+b_{22}+C_{I}}{4},\\
 m&=&\frac{ap+d}{8},\\
 g&=&\frac{p(1-p)(ap+b)}{2},\\
 k&=&\frac{ap(1-p)}{4},
 \EEA
 and $p$ is the probability of selecting the first action, which is the same for all the agents in the symmetric network configuration.  
The six eigenvalues that determine the stability of the configuration can be calculated analytically and are as follows: 
\begin{eqnarray*}
 \lambda_{1}&=&2 k -T,\\
\lambda_{2}&=&-T- 2 v,\\
\lambda_{3,4}&=&\frac{1}{2}(-k- 2 T +v- \sqrt{12 g m+(k+v)^2}),\\
\lambda_{5,6}&=&\frac{1}{2}(-k- 2 T +v+\sqrt{12 g m+(k+v)^2}).
\end{eqnarray*}
These expressions can be used to (numerically) identify the stability region of the configuration in the parameter space $(T,C_I)$, as shown in Fig.~\ref{fig9}.

\end{document}